\documentclass{PoS}
\usepackage{amsmath,amssymb}

\title{Gauge-Higgs Grand Unification}

\ShortTitle{Gauge-Higgs Grand Unification}

\author{\speaker{Yutaka Hosotani}\\%
       Department of Physics, Osaka University, Japan\\
       E-mail: \email{hosotani@phys.sci.osaka-u.ac.jp}}

\author{Naoki Yamatsu\\
        Department of Physics, Osaka University, Japan\\
        E-mail: \email{yamatsu@het.phys.sci.osaka-u.ac.jp}}

\abstract{Gauge-Higgs grand unification is formulated.  By extending $SO(5) \times U(1)_X$
gauge-Higgs electroweak unification, strong interactions are incorporated in $SO(11)$
gauge-Higgs unification in the Randall-Sundrum warped space.
Quarks and leptons are contained in spinor and vector  multiplets  of $SO(11)$.
Although the KK scale can be as low as $10\,$TeV, proton decay 
is forbidden by a conserved fermion number in the absence of Majorana masses of neutrinos. 
(OU-HET 880)
}

\FullConference{18th International Conference From the Planck Scale to the Electroweak Scale \\
		25-29 May 2015\\
		Ioannina, Greece }

\def\ignore#1{{}}

\newcommand{\beeq}{\begin{equation}}
\newcommand{\eneq}{\end{equation}}
\newcommand{\beqn}{\begin{eqnarray}}
\newcommand{\eeqn}{\end{eqnarray}}

\newcommand{\siml}{%
\hspace{0.3em}\raisebox{0.4ex}{$<$}\hspace{-0.75em}\raisebox{-.7ex}{$\sim$}\hspace{0.3em}}

\def\dd{\partial}
\def\la{\raise.16ex\hbox{$\langle$}\lower.16ex\hbox{}  }
\def\ra{\raise.16ex\hbox{$\rangle$}\lower.16ex\hbox{} }
\def\go{\rightarrow}

\def\onehalf{ \hbox{$\frac{1}{2}$} }

\def\onethird{ \hbox{$\frac{1}{3}$} }

\def\eff{{\rm eff}}

\def\cD{{\cal D}}

\def\SM{{\rm SM}}
\def\EM{{\rm EM}}
\def\EW{{\rm EW}}
\def\diag{{\rm diag ~}}
\def\GUT{{\rm GUT}}

\def\KK{{\rm KK}}

\begin{document}

\section{Gauge-Higgs unification}

We are in search of a principle for the 125 GeV Higgs scalar boson, 
which regulates Higgs couplings, explains how the electroweak (EW) gauge symmetry breaking 
takes place, and solves the gauge-hierarchy problem.
One possible, promising answer  is the gauge-Higgs unification.

%The 4D Higgs boson is identified as a part of the gauge fields in extra dimensions. 
In the gauge-Higgs EW unification one starts with gauge theory, say, in 5 dimensions.
4-dimensional components of gauge potential $A_M$ contain 4D gauge fields such as
photon $\gamma$, $W$ and $Z$ bosons.  The extra dimensional component 
transforms as a scalar under 4-dimensional Lorentz transformations, and its zero mode
is identified with the 4D Higgs scalar field.
When the extra dimensional space is not simply connected, there arises
an Aharonov-Bohm (AB) phase $\theta_H$ along the extra dimension.  The 4D Higgs field
is a 4D fluctuation mode of the AB phase. 
The value of $\theta_H$ is not determined at the classical level, but is dynamically 
determined at the quantum level.  
In non-Abelian gauge theory it may lead to spontaneous gauge symmetry breaking.
It is called the Hosotani mechanism.\cite{YH1}-\cite{HHHK2003}

The most notable feature of gauge-Higgs unification by the Hosotani mechanism is 
that the finite mass of the Higgs boson is generated quantum mechanically, 
independent of a cutoff scale.   Further the interactions of the Higgs boson with
itself and other fields are governed by the gauge principle so that  the model is 
very restrictive and predictive.  

\section{$SO(5) \times U(1)$  EW unification}

There is a realistic model of gauge-Higgs EW unification.  It is the $SO(5) \times U(1)_X$ gauge-Higgs 
EW unification in the Randall-Sundrum (RS) warped space.\cite{ACP2005}-\cite{HZgamma}
The metric of the RS space is given by 
$ds^2 = e^{-2 \sigma(y)} \eta_{\mu\nu} dx^\mu dx^\nu + dy^2$, where
$\eta_{\mu\nu} = \diag (-1, 1, 1, 1)$, 
$\sigma (y) = \sigma(-y) = \sigma(y+ 2L)$, and $\sigma(y) = k y$ for $0 \le y \le L$.
$z_L = e^{kL} \gg 1$ is called the warp factor.   
The bulk part $0 < y < L$ is an AdS space with a cosmological constant $\Lambda = - 6k^2$, 
which is  sandwiched by the Planck brane  at $y=0$ and  the TeV brane at $y=L$.
The RS is an orbifold; spacetime points $(x^\mu, y)$, $(x^\mu, -y)$ and $(x^\mu, y+ 2L)$
are identified. 

$SO(5) \times U(1)_X$ gauge theory is defined in RS.  Although physical quantities must be single-valued,
the gauge potential $A_M(x,y)$ may not be.  It satisfies
\beeq
\begin{pmatrix} A_\mu \cr A_y \end{pmatrix} (x , y_j -y) = 
P_j \begin{pmatrix} A_\mu \cr - A_y \end{pmatrix} (x , y_j + y) P_j^{-1} ~, 
 (y_0, y_1) = (0, L)
\label{BC1}
\eneq
where $P_j \in SO(5)$ up to sign and $P_j^2  = 1$.   Note that 
$A_M ( x, y+ 2L) = U A_M ( x, y) U^{-1}$ where $U = P_1 P_0$.
The set $\{ P_0, P_1 \}$ defines orbifold boundary conditions.  One chooses
\beeq
P_0 = P_1 = \begin{pmatrix} I_{4} & \cr & - 1  \end{pmatrix} ~,
%P_0 = P_1 = \diag (1, 1, 1,1, -1) ~,
\label{BC2}
\eneq
which breaks $SO(5)$ to $SO(4) \simeq SU(2)_L \times SU(2)_R$.
Quark- and lepton-multiplets are introduced in the bulk in the vector 
representation of $SO(5)$, whereas dark fermion multiplets in the spinor representation.
In addition to them, brane scalar field $\Phi$ and brane fermions fields are introduced
on the Planck brane.  The brane scalar field $\Phi (x)$ spontaneously breaks
$SU(2)_R \times U(1)_X$ to $U(1)_Y$, reducing the original symmetry $SO(5) \times U(1)_X$
to the standard model (SM) symmetry $SU(2)_L \times U(1)_Y$.

The zero modes of $A_y (x,y)$ reside in the $SO(5)/SO(4)$ part, $A_y^{(a 5)}$ ($a=1 \sim 4$)
in the standard notation.  They transform as an $SO(4)$ vector, or an $SU(2)_L$ doublet.
Three out of the four components are absorbed by $W$ and $Z$ gauge bosons.  
The unabsorbed component becomes the neutral Higgs boson $H(x)$ of mass 125 GeV.  
The Higgs boson appears as an AB phase;
\beqn
&&\hskip -1.cm
e^{i \hat \theta (x) } = P \exp \Big\{ ig_A \int_0^{2L} dy \, A_y (x, y) \Big\} ~, \cr
\noalign{\kern 10pt}
&&\hskip -1.cm
\hat \theta (x)  = \theta_H + \frac{H(x)}{f_H} ~~, ~~~
f_H = \frac{2}{g_A} \sqrt{\frac{k}{z_L^2 -1}} 
\sim \frac{2 m_\KK}{\pi g_w \sqrt{kL}}  ~.
\label{AB1}
\eeqn
Here $g_w$ is the 4D $SU(2)_L$ gauge coupling constant.  
The KK mass scale is given by $m_\KK = \pi k /(z_L -1)$.
The gauge invariance guarantees that physics is periodic in $\theta_H$ with a period $2\pi$.

The model is successful, being consistent with data at low energies for $\theta_H \siml 0.1$.
The values of the parameters of the model are determined such that the observed $m_Z$, 
$g_w$, $\sin \theta_W$, $m_H$ and quark/lepton masses are reproduced.
There remain two relevant free parameters, $z_L$ and $n_F$ (the number of dark fermion
multiplets).  With $z_L$ and $n_F$ given, the effective potential $V_\eff (\theta_H)$
is evaluated.  The value of $\theta_H$ is determined dynamically by the location of 
the global minimum of $V_\eff (\theta_H)$.

\begin{figure}[tbh]\begin{center}
{\includegraphics*[width=.55\linewidth]{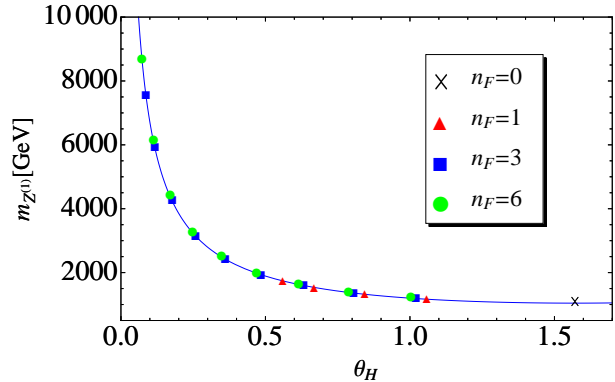}}
\caption{
$\theta_H$ vs $m_{Z^{(1)}}$ for $m_H=126 \,$GeV with $n_F$ degenerate dark fermions.
}
\label{thetaHZ1}
\end{center}
\end{figure}

Quite remakable is the fact that most of the important physical quantities such as
the Higgs couplings, the KK  scale $m_\KK$, and the KK spectrum of gauge bosons and quarks/leptons
are approximately determined as functions of $\theta_H$, independent of detailed values  of $(z_L, n_F)$.
There hold universality relations.\cite{FHHOS2013, LHCsignalsDM}
For instance, the mass $m_{Z^{(1)}}$ of the first KK $Z$, depicted in Fig.\ \ref{thetaHZ1}, and the KK 
scale $m_\KK$ satisfy 
\beqn
&&\hskip -1.cm
m_{Z^{(1)}} \sim 1044 \, {\rm GeV} ~ {(\sin \theta_H)^{-0.808}} ~, \cr
\noalign{\kern 5pt}
&&\hskip -1.cm
m_\KK \sim  {1352 \, {\rm GeV}} ~ {(\sin \theta_H)^{-0.786}} ~.
\label{mass1}
\eeqn

%%%%%%%%%%%%%%%%%%%%
%%%%%%%%%%%%%%%%%%%%

Similarly the Higgs cubic and quartic self-couplings are given by 
\beqn
&&\hskip -1.cm
\lambda_3 / {\rm GeV} \sim  26.7 \cos \theta_H + 1.42 (1+ \cos 2\theta_H)  ~, \cr
\noalign{\kern 10pt}
&&\hskip -1.cm
\lambda_4 \sim  - 0.0106 + 0.0304 \cos 2 \theta_H + 0.00159 \cos 4 \theta_H ~.
\label{HiggsCoupling}
\eeqn
These numbers should be compared with $\lambda_3^\SM = 31.5\,$GeV and 
$\lambda_4^\SM= 0.0320$ in SM.

\begin{figure}[ht]\begin{center}
{\includegraphics*[width=.55\linewidth]{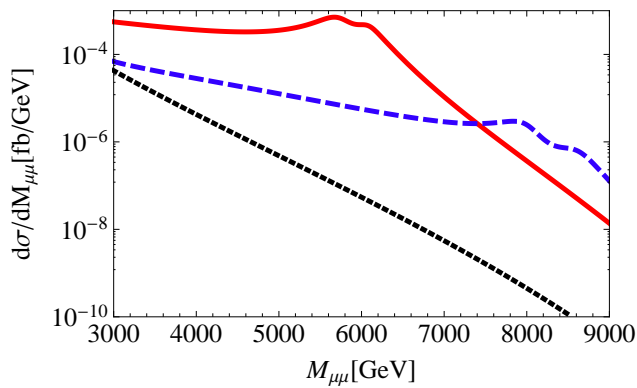}}
\caption{
The differential cross section for $pp \to \mu^+ \mu^- X$ 
at the 14 TeV LHC for $\theta_H=0.114$ (red solid curve) and 
for $\theta_H=0.073$ (blue dashed curve) .
The nearly straight black line represents the SM background.
}
\label{Zprime}
\end{center}
\end{figure}

The model gives definitive prediction for $Z'$ events at LHC.
The $e^+ e^-$ or $\mu^+ \mu^-$ signals through virtual production of 
$\gamma^{(1)},  Z^{(1)},  Z_R^{(1)}$ should be detected at the 14 TeV LHC.
The predicted cross section is shown in Fig.\ \ref{Zprime}.
The widths are large in the gauge-Higgs unification, as the gauge couplings of the first
KK modes are large for right-handed quarks and leptons.

%%%%%%%%%%%%%%%%%%%%%%%%%%%%%%%%%%%%%%%%%%%%%%%
%%%%%%%%%%%%%%%%%%%%%%%%%%%%%%%%%%%%%%%%%%%%%%%

\section{$SO(11)$ grand unification}

What is next?  It is most natural to extend the gauge-Higgs unification scenario
to incorporate strong interactions.  We would like to have a theory with  gague group ${\cal G}$,
which reduces to the SM symmetry $SU(3)_C \times SU(2)_L \times U(1)_Y$ at low energies, 
and in which 4D Higgs boson of $m_H = 125 \,$GeV appears as a part of the 
extra-dimensional component of gauge fields.  There have been several attempts along 
this line.    $SU(6)$ gauge-Higgs unification was considered on 
$M^4 \times (S^1/Z_2)$.\cite{Burdman2003, HHKY2004, Lim2007}
However it necessarily yields an extra $U(1)$, and  dynamical EW symmetry
breaking can be achieved only with extra matter fields resulting in exotic particles at low
energies.  $SU(5) \times SU(5)$ unification model has been proposed.\cite{Kojima2011}
There is an approach from the composite Higgs scenario.\cite{Serra2011}
None of these models is satisfactory  with the phenomenology below the EW scale.

$SO(5) \times U(1)_X$ gauge-Higgs unification in RS space is a good, realistic model
at low energies.  One might expect that in gauge-Higgs grand unification
the gauge group ${\cal G}$ reduces to $SU(3)_C\times SO(5) \times U(1)_X$ at
some scale, and further to $SU(3)_C\times SU(2)_L \times U(1)_Y$ at a lower scale.
This approach, however,  turns out  not to work.

We propose $SO(11)$ gauge-Higgs grand unification in the RS space.\cite{HY2015}
In the bulk region $(0 \le y \le L)$ there are, in addition to the $SO(11)$ gauge fields $A_M$,
fermion multiplets in the $SO(11)$ spinor representation, $\Psi_{\bf 32}$, and 
in the $SO(11)$ vector representation, $\Psi_{\bf 11}$.  On the Planck brane $(y=0)$ 
a scalar field in the $SO(10)$ spinor representation, $\Phi_{\bf 16}$, is introduced.
The symmetry breaking pattern in this scenario is
\beqn
SO(11) &\go& SO(4) \times SO(6)  
\simeq SU(2)_L \times SU(2)_R \times  SU(4) \quad \hbox{by BC} \cr
\noalign{\kern 10pt}
&\go&  SU(3)_C \times SU(2)_L \times U(1)_Y \qquad \hbox{by } \la \Phi_{\bf 16} \ra \cr
\noalign{\kern 10pt}
&\go&  SU(3)_C \times U(1)_\EM  \qquad \hbox{by } \theta_H ~ .
\label{SymBreaking1}
\eeqn

The first step of the symmetry breaking in \eqref{SymBreaking1} is achieved by
the orbifold boundary condition, which is given, in the form of \eqref{BC1}, with
\beeq
P_0 = \begin{pmatrix} I_{10} & \cr & -1 \end{pmatrix} ~,~~~
P_1 = \begin{pmatrix} I_{4} & \cr & - I_7 \end{pmatrix} ~.
\label{BC5}
\eneq
$P_0$ and $P_1$ break $SO(11)$ to $SO(10)$ at the Planck brane and
to $SO(4) \times SO(7)$ at the TeV brane, respectively.  With these two combined
the symmetry is reduced to 
$SO(4) \times SO(6)  \simeq SU(2)_L \times SU(2)_R \times  SU(4)$.
At this stage $A_\mu(x,y)$ has zero modes (4D massless gauge fields) 
in $SO(4) \times SO(6)$, whereas $A_y(x,y)$ has zero modes only in
the $A_y^{(a \, 11)}$ ($a=1 \sim 4$) components, which become the 4D Higgs doublet.

On the Planck brane $SO(10)$ gauge invariance is maintained.  
The brane scalar field $\Phi_{\bf 16}$ spontaneously breaks $SO(10)$  
to $SU(5)$. The resultant symmetry is $SU(5) \cap [SO(4) \times SO(6)]$, namely
$SU(3)_C \times SU(2)_L \times U(1)_Y$.

The last step  in \eqref{SymBreaking1} is induced by the Hosotani mechanism.
The EW symmetry breaking takes place by dynamics of 
the AB phase $\theta_H$ associated with $A_y^{(4 \, 11)}$ in \eqref{AB1}.
In this scheme generators of $U(1)_Y$ and $U(1)_\EM$ are given, in terms of 
$SO(11)$ generators, by
\beqn
&&\hskip -1.cm
Q_Y = \onehalf ( T_{12} - T_{34}) - \onethird (T_{56} + T_{78} + T_{9\, 10}) ~, \cr
\noalign{\kern 10pt}
&&\hskip -1.cm
Q_\EM =  T_{12}  - \onethird (T_{56} + T_{78} + T_{9\, 10}) ~.
\label{EMcharge}
\eeqn
It follows that the Weinberg angle is the same as in the $SU(5)$ or $SO(10)$ GUT;
\beeq
g_Y' = \sqrt{\frac{3}{5}} \, g_w ~, ~~~ e = \sqrt{\frac{3}{8}} \, g_w ~, ~~~
\sin^2 \theta_W = \frac{3}{8} ~.
\label{Wangle}
\eneq

\section{Quarks and leptons}

We introduce fermions $\Psi_{\bf 32}$ and $\Psi_{\bf 11}$ in the bulk.
Quarks and leptons in SM are contained mostly in $\Psi_{\bf 32}$.  To see it explicitly,
we take the following representation of $SO(11)$ Clifford algebra 
$\{ \Gamma_j , \Gamma_k \} = 2 \delta_{jk} \, I_{32}$ ($j,k = 1 \sim 11$);
\beqn
&&\hskip -1.cm
\Gamma_{1,2,3} = \sigma^{1,2,3} \otimes \sigma^1 \otimes \sigma^1 \otimes \sigma^1 \otimes \sigma^1, \cr
&&\hskip -1.cm
\Gamma_{4,5} = \sigma^0 \otimes \sigma^{2,3} \otimes \sigma^1 \otimes \sigma^1 \otimes \sigma^1, \cr
&&\hskip -1.cm
\Gamma_{6,7} = \sigma^0 \otimes \sigma^0 \otimes \sigma^{2,3} \otimes \sigma^1 \otimes \sigma^1, \cr
&&\hskip -1.cm
\Gamma_{8,9} = \sigma^0 \otimes \sigma^0 \otimes \sigma^0 \otimes \sigma^{2,3} \otimes \sigma^1, \cr
&&\hskip -1.cm
\Gamma_{10,11} = \sigma^0 \otimes \sigma^0 \otimes \sigma^0 \otimes \sigma^0 \otimes \sigma^{2,3} ,
\label{Clifford1}
\eeqn
where  $\sigma^0 = I_2$ and $\sigma^{1,2,3}$ are Pauli matrices.  
The $SO(11)$ generators are given by
$T_{jk}^{\rm sp} = - \onehalf i  \Gamma_j \Gamma_k$ ($j \not= k$).  
The upper and lower half components of $\Psi_{\bf 32}$ 
correspond to ${\bf 16}$ and $\overline{\bf 16}$ of $SO(10)$.
The orbifold boundary condition matrices in the spinorial representation are given by
$P_0^{\rm sp} = \Gamma_{11}$ and 
$P_1^{\rm sp} = I_2 \otimes \sigma^3 \otimes I_8$. 
$\Psi_{\bf 32}$ and $\Psi_{\bf 11}$ satisfy
\beqn
&&\hskip -1.cm
\Psi_{\bf 32} (x, y_j -y) = 
-P_j^{\rm sp} \, \gamma^5 \, \Psi_{\bf 32} (x,y_j+y) ~, \cr
\noalign{\kern 5pt}
&&\hskip -1.cm
\Psi_{\bf 11} (x, y_j -y) = 
\eta^{\bf 11}_j  P_j \, \gamma^5 \, \Psi_{\bf 11} (x,y_j+y) ~,
\label{BC3}
\eeqn
where $\eta^{\bf 11}_j = +1$ or $-1$.  The bulk action for the fermions takes the form 
\beeq
\int d^5x \sqrt{- \det G} \Big\{ \overline{\Psi}_{\bf 32} \cD (c_{\bf 32}) \Psi_{\bf 32}
+ \overline{\Psi}_{\bf 11} \cD (c_{\bf 11}) \Psi_{\bf 11} \Big\} 
\label{action2}
\eneq
where $\cD (c) = \gamma^A {e_A}^M D_M - c \sigma '(y)$ and
$D_M = \dd_M + \frac{1}{8} \omega_{MBC} [\gamma^B, \gamma^C] - i g A_M$.
The generators of $SU(2)_L$ and $SU(2)_R$ are given by
\beeq
\Big[ T^a_L , T^a_R \Big] =  \onehalf \sigma^a \otimes 
\bigg[ \begin{pmatrix} 1\cr &0 \end{pmatrix} , \begin{pmatrix} 0\cr &1 \end{pmatrix}\bigg] \otimes I_8 ~.\label{generator1}
\eneq

The content of $\Psi_{\bf 32}$ is given by
\beqn
&&\hskip -1.cm
\Psi_{\bf 32} = \begin{pmatrix} \Psi_{\bf 16} \cr \Psi_{\overline{\bf 16}} \end{pmatrix} ~,~~ 
%\cr
%\noalign{\kern 10pt}
%&&\hskip -1.cm
\Psi_{\bf 16} = \begin{pmatrix} \ell \cr \hat q_1 \cr q_3 \cr \hat q_2 \cr
\noalign{\kern 5pt}
q_1 \cr \hat \ell \cr q_2 \cr \hat q_3  \end{pmatrix} ~, ~~
\Psi_{\overline{\bf 16}} = \begin{pmatrix} \hat q_3' \cr q_2' \cr \hat \ell' \cr q_1' \cr
\noalign{\kern 5pt}
\hat q_2' \cr q_3' \cr \hat q_1' \cr \ell'   \end{pmatrix} ~, ~~ \cr
\noalign{\kern 20pt}
&&\hskip -1.cm
\begin{matrix}
\ell = \begin{pmatrix} \nu \cr e \end{pmatrix} ,
& q_j = \begin{pmatrix} u_j \cr d_j \end{pmatrix}, \cr
\noalign{\kern 10pt}
\hat \ell = \begin{pmatrix} \hat e \cr \hat \nu \end{pmatrix},
& \hat q_j = \begin{pmatrix} \hat d_j \cr \hat u_j \end{pmatrix} ,
\end{matrix}
~~
\begin{matrix}
\ell' = \begin{pmatrix} \nu' \cr e' \end{pmatrix} ,
& q_j' = \begin{pmatrix} u_j' \cr d_j '\end{pmatrix}, \cr
\noalign{\kern 10pt}
\hat \ell' = \begin{pmatrix} \hat e' \cr \hat \nu' \end{pmatrix},
& \hat q_j' = \begin{pmatrix} \hat d_j' \cr \hat u_j' \end{pmatrix} ,
\end{matrix}  
\label{content1}
\eeqn
A field with hat has an opposite charge to the corresponding one without hat.
$u_j$ $(u_j')$ and $\hat u_j$ $(\hat u_j')$, for instance,  have 
$Q_\EM = + \frac{2}{3}$ and $- \frac{2}{3}$, respectively.
With the orbifold boundary condition \eqref{BC3}, zero modes of $\Psi_{\bf 32}$
appear in
\beeq
%\hbox{zero modes :} ~ 
\ell_L = \begin{pmatrix} \nu_L \cr e_L \end{pmatrix} ~,~~
q_{jL} = \begin{pmatrix} u_{jL} \cr d_{jL} \end{pmatrix} ~,~~
\ell'_R = \begin{pmatrix} \nu'_R \cr e'_R \end{pmatrix} ~,~~
q_{jR}' = \begin{pmatrix} u_{jR}' \cr d_{jR}'\end{pmatrix} ~.
\label{zeroM1}
\eneq
All of the SM fermions, but nothing else, appear in $\Psi_{\bf 32}$ as zero modes.  

The content of $\Psi_{\bf 11}$ is 
\beeq
\Psi_{\bf 11} = \Bigg[ \begin{pmatrix} \hat E & N \cr \hat N  & E \end{pmatrix} ;
\big(D_j , \hat D_j \big) ; ~ S  ~\Bigg] ~.
\label{content2}
\eneq
$N$,  $E$,  and $D_j$ have the same electric charges as $\nu$,  $e$,  and $d_j$, respectively.
$S$ is an $SO(10)$ singlet, and is neutral.  With given $(\eta^{\bf 11}_0, \eta^{\bf 11}_1)$
in the boundary condition \eqref{BC3}, zero modes are found in
\beeq
\begin{matrix}
(+,+) : 
&\begin{pmatrix} \hat E_R & N_R \cr \hat N_R  & E_R \end{pmatrix} ,~~ S_L
&~~~~~(-,-) : 
&\begin{pmatrix} \hat E_L & N_L \cr \hat N_L  & E_L \end{pmatrix} ,~~ S_R \cr
\noalign{\kern 10pt}
(+,-) :
&D_{jR} , \hat D_{jR} 
&~~~~~(-,+) : 
&D_{jL} , \hat D_{jL} 
\end{matrix}
\eneq

\section{KK spectrum}
To see whether or not the EW symmetry is dynamically broken, one need to know
all KK mass spectra which depend on $\theta_H$.  In the gauge field sector
the $W$ tower, the $Z$ tower, and $Y$ boson tower have $\theta_H$-dependent
spectra.  The spectra are found from zeros of several equations given by
\beqn
%&&\hskip -1.cm
W ~\hbox{tower:} &&  2 S(1;\lambda_n) C'(1; \lambda_n) + \lambda_n \sin^2 \theta_H = 0 ~, \cr
\noalign{\kern 5pt}
%&&\hskip -1.cm
Z ~~\hbox{tower:} &&5 S(1;\lambda_n) C'(1; \lambda_n) + 4 \lambda_n \sin^2 \theta_H = 0 ~, \cr
\noalign{\kern 5pt}
Y ~~\hbox{tower:} && 2 S(1;\lambda_n) C'(1; \lambda_n) + \lambda_n (1 + \cos^2 \theta_H) = 0 ~.
\label{spectrum1}
\eeqn
Here $C(z;\lambda) = \onehalf \pi \lambda z z_L F_{1,0} (\lambda z, \lambda z_L)$
and $S(z;\lambda) = -\onehalf \pi \lambda z  F_{1,1} (\lambda z, \lambda z_L)$
where $F_{\alpha,\beta} (u,v) = J_\alpha (u) Y_\beta (v) - Y_\alpha (u) J_\beta (v)$.
$z= e^{ky}$ and $C' = dC /dz$.  The mass is given by $m_n= k\lambda_n$.  
The lowest modes of $W$ and $Z$ towers are $W$ and $Z$ bosons.  Their masses are found to be
\beeq
m_W \sim \frac{\sin \theta_H}{\pi \sqrt{kL}} ~ m_\KK ~, ~~
m_Z \sim \frac{m_W}{\cos \theta_W} ~, ~~
\sin^2 \theta_W = \frac{3}{8} ~.
\eneq
Among $A_y$, the components $[\tilde A_y^{a4}, \tilde A_y^{a11} ]$  ($a= 1\sim 3, 5 \sim 10$) 
have $\theta_H$-dependent spectra given by
\beeq
S(1;\lambda_n) C'(1; \lambda_n) + \lambda_n  
\begin{pmatrix}\sin^2 \theta_H \cr \cos^2 \theta_H \end{pmatrix} = 0 \quad
\hbox{for~} a = \Big\{ \,  \begin{matrix}  1 \sim 3 ~,    \cr  5 \sim 10 . \end{matrix} 
\label{Sspectrum}
\eneq

 In the absence of brane interactions the spectrum of the $\Psi_{\bf 32}$ tower is found to be
 \beeq
 S_L (1; \lambda_n , c_{\bf 32}) S_R (1; \lambda_n , c_{\bf 32}) 
 +\begin{pmatrix} \sin^2 \onehalf \theta_H \cr
 \noalign{\kern 5pt} \cos^2 \onehalf \theta_H \end{pmatrix} = 0 
\label{SpFspectrum}
\eneq
where the upper component is 
for $\ell, \ell', q_j, q_j'$ and the lower component 
for $\hat \ell, \hat \ell', \hat q_j, \hat q_j'$.
Here $S_{L/R} (z; \lambda, c) = \mp \onehalf \pi \lambda \sqrt{z z_L} 
F_{c \pm \onehalf, c \pm \onehalf} (\lambda z, \lambda z_L)$.  
For $\Psi_{\bf 11}$ the 4th and  11th components mix, and their spectrum is given by
\beeq
 S_L (1; \lambda_n , c_{\bf 11}) S_R (1; \lambda_n , c_{\bf 11}) + 
 \begin{pmatrix} \sin^2  \theta_H \cr \cos^2  \theta_H \end{pmatrix} = 0 
\label{VecFspectrum}
\eneq
for $\eta^{\bf 11}_ 0 \eta^{\bf 11}_ 1 = \pm 1$.
To get the observed quark/lepton spectrum, one must take account of brane interactions
among $\Psi_{\bf 32}$, $\Psi_{\bf 11}$ and $\Phi_{\bf 16}$.

\section{EW symmetry breaking}

One need to evaluate $V_\eff (\theta_H)$
to find if the EW symmetry breaking takes place by the Hosotani mechanism.
Full analysis must be waited for until parameters
in the brane interactions  are fixed to reproduce the observed quark-lepton spectrum.
Here we point out that even in the absence of fermions the EW symmetry breaking 
occurs in the $SO(11)$ gauge-Higgs unification.

\begin{figure}[htb]
\centering
\includegraphics[width=8cm]{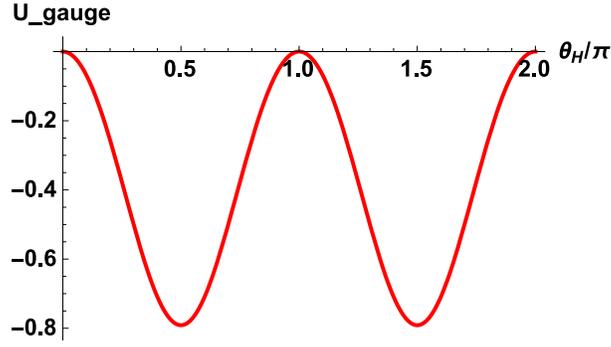}
\caption{
$U = (4\pi)^2 (kz_L^{-1})^{-4} V_\eff  (\theta_H)$  in pure gauge theory
 is plotted  in the $\xi=0$ gauge. 
$V_\eff (\theta_H)$ with a minimum at $0< \theta_H < \onehalf \pi$
is achieved with the inclusion of fermions and brane interactions.
}
\label{eff-potential1}
\end{figure}

In pure gauge theory $V_\eff (\theta_H)$ is evaluated with 
\eqref{spectrum1} and \eqref{Sspectrum}.  See Fig.\ \ref{eff-potential1}.
It  has the global minimum at $\theta_H = \pm \onehalf \pi$, and 
the EW gauge symmetry is dynamically broken.
This has never happened in the gauge-Higgs EW unification models.
The symmetry breaking is caused, because 
in the current model there are six $Y$ towers with the spectrum in \eqref{spectrum1}
where the lowest modes have the smallest mass for $\cos \theta_H=0$.

The minimum at $\theta_H = \onehalf \pi$, however, is not acceptable phenomenologically, 
as it leads to a stable Higgs boson due to the $H$ parity.\cite{HKT2009, HTU2011}
Desirable value of $\theta_H \siml 0.1$ can be achieved by including fermion
multiplets $\Psi_{\bf 32}$ and $\Psi_{\bf 11}$ and brane interactions.

\section{Energy scales in gauge-Higgs grand unification}

There are several energy scales in the gauge-Higgs grand unification in RS.
\beqn
&&\hskip -1.cm
(1) ~ \hbox{Size of the 5th dimension: } E_{\rm size} = \frac{\pi}{L} \cr
\noalign{\kern 5pt}
&&\hskip -1.cm
(2) ~ \hbox{GUT scale: } m_\GUT \cr
\noalign{\kern 5pt}
&&\hskip -1.cm
(3) ~ \hbox{KK scale: }  
m_\KK = \frac{\pi k}{z_L -1} \sim \pi k e^{- kL} 
\sim \frac{\sqrt{kL}}{\sin \theta_H} ~ m_W \cr
\noalign{\kern 5pt}
&&\hskip -1.cm
(4) ~ \hbox{EW scale: }  m_\EW \cr
\noalign{\kern 5pt}
&&\hskip -1.cm
(5) ~ \hbox{QCD scale: }  \Lambda_{QCD} 
\label{scale1}
\eeqn
In the flat $M^4 \times (S^1/Z_2)$, $E_{\rm size} = m_\KK = 1/R$,
but in the RS space, $E_{\rm size} \gg m_\KK \gg m_\EW$.

The GUT scale $m_\GUT$ is defined by the gauge coupling unification.
In the gauge-Higgs unification  KK modes of gauge fields and fermions 
are excited above $m_\KK$ in GUT multiplets, which does not necessarily change 
the GUT unification scale so much.  The preliminary study indicates
that $m_\KK \ll m_\GUT$.  It is possible to have $m_\GUT \sim E_{\rm size}$.

\section{Forbidden proton decay}

If KK excited states of $X$ and $Y$ bosons show up above $m_\KK$,
one may worry about large decay rate of protons.  In 4D $SU(5)$ or $SO(10)$ GUT, 
for instance, proton decay proceeds through $X$ and $Y$ boson exchange.  
The masses of $X$ and $Y$ bosons are $O(m_\GUT)$ where $m_\GUT \sim 10^{15}\,$GeV.
In the gauge-Higgs unification $m_\KK$ is much lower.  For $\theta_H \sim 0.1$,
$m_\KK \sim 10\,$TeV so that  $X$ and $Y$ boson exchange may lead to 
rapidly decaying protons.

Remarkably the proton decay is forbidden in the $SO(11)$ gauge-Higgs grand unification,
provided Majorana mass terms for neutrinos are absent.
Notice that all quarks and leptons are contained in $\ell, q_j. \ell', q_j'$ in 
$\Psi_{\bf 32}$ in \eqref{content1} and \eqref{zeroM1}.
All of them have a fermion number $N_{\Psi} = +1$.
In the presence of brane interactions on the Planck brane,   $\Psi_{\bf 32}$ 
and $\Psi_{\bf 11}$ mix, but still the fermion number $N_{\Psi} $ is conserved.
Proton has $N_\Psi = +3$, which cannot decay to, say,  $e^+ \pi^0$ that
has $N_\Psi = -1$.

This should be contrasted to 4D GUT.  In 4D $SU(5)$ GUT,  
$\Psi_{\overline{\bf 5}}$ and $\Psi_{\bf 10}$ contains $(\ell_L, d^c_{jL})$ and
$(q_{jL}, u^c_{jL}, e^c_L)$, respectively  so that gauge interactions 
lead, for instance,  to $u \go u^c + X$ and $d + X \go e^+$, which results
in $uud \go u u^c e^+$, namely proton decay $p \go \pi^0 e^+$.
Such transitions do not take place in the $SO(11)$ gauge-Higgs grand unification
as a consequence of the $N_\Psi$ conservation.
In 4D $SO(10)$ GUT, $\Psi_{{\bf 16}\,L}$ contains all quarks and leptons, which are obtained
from the $\Psi_{{\bf 16}}$ content in \eqref{content1} by replacing
$\ell, q_j, \hat \ell, \hat q_j$ by $\ell_L, q_{jL},   \ell^c_L, q^c_{jL}$.
Consequently gauge interactions induce proton decay as in 4D $SU(5)$ GUT.
In the $SO(11)$ theory the zero modes $\ell^c_L, q^c_{jL}$ reside in the $\Psi_{\overline{\bf 16}}$
part of $\Psi_{\bf 32}$.  In the $SO(11)$ gauge-Higgs unification, the number of 
components of spinor representation is doubled, compared to that in $SO(10)$ theory, 
from 16 to 32, but the orbifold boundary conditions reduce the number of chiral zero modes
to 16.

If Majorana mass terms were introduced for neutrinos on the Planck brane, 
the $N_\Psi$ fermion number would not be conserved.  It would give rise to proton 
decay at the higher loop level.  Its rate would be suppressed if Majorana masses were
sufficiently large.

\section{Summary}

Grand unification is necessary to explain the observed charge quantization in quarks 
and leptons.  We have formulated the $SO(11)$ gauge-Higgs grand unification
in which the 4D Higgs boson of $m_H=125\,$GeV appears as a part of gauge fields
in 5 dimensions.  It generalizes the $SO(5) \times U(1)_X$ gauge-Higgs EW unification.
Dynamical EW gauge symmetry breaking is achieved by the Hosotani mechanism.
The $SO(11)$ structure appears above $m_\KK$, which can be as low as 10$\,$TeV.
Nevertheless the stability of protons is guaranteed by the conservation of 
the new fermion number $N_\Psi$.

There remain many problems to be clarified.  First of all, one has to determine
the parameters of the model, including brane interactions, such that the observed
mass spectrum of quarks and lepton is reproduced, and the EW symmetry breaking
is indeed spontaneously broken.
Gauge coupling unification need to  be examined by solving RGE.
The scenario of the gauge-Higgs unification is promising.

\vskip 20pt

\subsection*{Acknowledgements}

%\noindent 
%{\bf Acknowledgements}
This work was supported in part  by  JSPS KAKENHI grants No.\ 23104009 
and No.\ 15K05052.

%\end{document}
\newpage

% A useful Journal macro
%\def\jnl#1#2#3#4{{#1}{\bf #2} (#4) #3}
\def\jnl#1#2#3#4{{#1}{\bf #2},  #3 (#4)}

\def\Zphys{{\em Z.\ Phys.} }
\def\jssc{{\em J.\ Solid State Chem.\ }}
\def\jpsJ{{\em J.\ Phys.\ Soc.\ Japan }}
\def\ptps{{\em Prog.\ Theoret.\ Phys.\ Suppl.\ }}
\def\PTP{{\em Prog.\ Theoret.\ Phys.\  }}
\def\PTEP{{\em Prog.\ Theoret.\ Exp.\  Phys.\  }}
\def\JMP{{\em J. Math.\ Phys.} }
\def\NPB{{\em Nucl.\ Phys.} B}
\def\NP{{\em Nucl.\ Phys.} }
\def\PLB{{\it Phys.\ Lett.} B}
\def\PL{{\em Phys.\ Lett.} }
\def\PRL{\em Phys.\ Rev.\ Lett. }
\def\PRB{{\em Phys.\ Rev.} B}
\def\PRD{{\em Phys.\ Rev.} D}
\def\PRe{{\em Phys.\ Rep.} }
\def\AP{{\em Ann.\ Phys.\ (N.Y.)} }
\def\RMP{{\em Rev.\ Mod.\ Phys.} }
\def\ZPC{{\em Z.\ Phys.} C}
\def\SCI{\em Science}
\def\CMP{\em Comm.\ Math.\ Phys. }
\def\MPLA{{\em Mod.\ Phys.\ Lett.} A}
\def\IJMPA{{\em Int.\ J.\ Mod.\ Phys.} A}
\def\IJMPB{{\em Int.\ J.\ Mod.\ Phys.} B}
\def\EPJC{{\em Eur.\ Phys.\ J.} C}
\def\PR{{\em Phys.\ Rev.} }
\def\JHEP{{\em JHEP} }
\def\JCAP{{\em JCAP} }
\def\cmp{{\em Com.\ Math.\ Phys.}}
\def\JPA{{\em J.\  Phys.} A}
\def\JPG{{\em J.\  Phys.} G}
\def\NJP{{\em New.\ J.\  Phys.} }
\def\CQG{\em Class.\ Quant.\ Grav. }
\def\ATMP{{\em Adv.\ Theoret.\ Math.\ Phys.} }
\def\ibid{{\em ibid.} }

\renewenvironment{thebibliography}[1]
         {\begin{list}{[$\,$\arabic{enumi}$\,$]}  % {\arabic{enumi}.}
         {\usecounter{enumi}\setlength{\parsep}{0pt}
          \setlength{\itemsep}{0pt}  \renewcommand{\baselinestretch}{1.0}
          \settowidth
         {\labelwidth}{#1 ~ ~}\sloppy}}{\end{list}}

%%%%  title of reference  %%%%
%\def\reftitle#1{{\it ``#1,''}}    %to print.
\def\reftitle#1{}                %to hide.

%%%%%%%%%%%%% BIBLIOGRAPHY (US) %%%%%%%%%%%%%%%%%%%%


\begin{thebibliography}{99}
%%%%%%%%%%%%%%%%%%%%%%%%%%%%%%%%%%%%%%%%%%%%%%%

\leftline{\large \bf References}


\bibitem{YH1}
Y.~Hosotani,
%\reftitle{Dynamical Mass Generation by Compact Extra Dimensions}
\jnl{\PLB}{126}{309}{1983};
%\reftitle{Dynamics of Nonintegrable Phases and Gauge Symmetry Breaking}
\jnl{\AP}{190}{233}{1989}.

\bibitem{Davies1}
  A.~T.~Davies and A.~McLachlan,
%\reftitle{Gauge group breaking by Wilson loops}
\jnl{\PLB}{200}{305}{1988};
%\reftitle{Congruency class effects in the Hosotani model}
\jnl{\NPB}{317}{237}{1989}.


\bibitem{Hatanaka1998}
H.\ Hatanaka, T.\ Inami and C.S.\ Lim,
%\reftitle{The gauge hierarchy problem and higher dimensional gauge theories}
\jnl{\MPLA}{13}{2601}{1998}.

\bibitem{Kubo2002}
M.\ Kubo, C.S.\ Lim and H.\ Yamashita,
%\reftitle{The Hosotani Mechanism in Bulk Gauge Theories with an Orbifold Extra   Space $S^1/Z_2$}
\jnl{\MPLA}{17}{2249}{2002}.

\bibitem{HHHK2003}
N.\ Haba, Masatomi Harada, Y.\ Hosotani and Y.\ Kawamura,
%\reftitle{Dynamical rearrangement of gauge symmetry on the orbifold $S^1/Z_2$}
\jnl{\NPB}{657}{169}{2003}; 
{\it Erratum}-\jnl{\ibid}{{\rm B}669}{381}{2003}.


\bibitem{ACP2005}
K.~Agashe, R.~Contino and A.~Pomarol,
%\reftitle{The Minimal Composite Higgs Model}
\jnl{\NPB}{719}{165}{2005}.

\bibitem{AgasheContino2006}
K.~Agashe and R.~Contino,
%\reftitle{The minimal composite Higgs model and electroweak precision tests}
\jnl{\NPB}{742}{59}{2006}.

\bibitem{SH1} 
Y.~Sakamura and Y.~Hosotani,
%``WWZ, WWH, and ZZH Couplings in the Dynamical Gauge-Higgs Unification in the
%Warped Spacetime,''
\jnl{\PLB}{645}{442}{2007}.


\bibitem{Sakamura1}
Y.~Sakamura,
%\reftitle{Effective theories of gauge-Higgs unification models in warped spacetime}
\jnl{\PRD}{76}{065002}{2007}.

\bibitem{MSW}
A.~D.~Medina, N.~R.~Shah and C.~E.~M.~Wagner, 
%\reftitle{Gauge-Higgs Unification and Radiative Electroweak Symmetry Breaking in
%Warped Extra Dimensions}
\jnl{\PRD}{76}{095010}{2007}.

\bibitem{Giudice2007}
G.F. Giudice,  C. Grojean,  A. Pomarol,  R. Rattazzi, 
\reftitle{The Strongly-Interacting Light Higgs}
\jnl{\JHEP}{0706}{045}{2007}.

\bibitem{HS2007} 
Y.~Hosotani and Y.~Sakamura,
%\reftitle{Anomalous Higgs Couplings in the $SO(5)\times U(1)_{B-L}$ Gauge-Higgs
%Unification in Warped Spacetime}
\jnl{\PTP}{118}{935}{2007}.


\bibitem{HOOS2008} 
Y.~Hosotani, K.~Oda, T.~Ohnuma and Y.~Sakamura,
%\reftitle{Dynamical Electroweak Symmetry Breaking in $SO(5) \times U(1)$
%Gauge-Higgs Unification with Top and Bottom Quarks}
\jnl{\PRD}{78}{096002}{2008}; 
{\it Erratum}-\jnl{\ibid}{{\rm D}79}{079902}{2009}.
% [arXiv:0806.0480 [hep-ph]].

\bibitem{HK}
Y.~Hosotani and Y.~Kobayashi,
%\reftitle{Yukawa Couplings and Effective Interactions in Gauge-Higgs Unification}
\jnl{\PLB}{674}{192}{2009}.

\bibitem{Pomarol2009}
B. Gripaios, A. Pomarol,  F. Riva,  J. Serra,
%\reftitle{Beyond the Minimal Composite Higgs Model}
\jnl{\JHEP}{0904}{070}{2009}.

\bibitem{HKT2009}
Y.~Hosotani, P.~Ko and M.~Tanaka,
%\reftitle{Stable Higgs Bosons as Cold Dark Matter}
\jnl{\PLB}{680}{179}{2009}.

\bibitem{HNU}
Y.\ Hosotani, S.\ Noda and N.\ Uekusa,
%\reftitle{The Electroweak gauge couplings in SO(5) x U(1) gauge-Higgs unification}
\jnl{\PTP}{123}{757}{2010}.

\bibitem{HTU2011}
Y.\ Hosotani, M.\ Tanaka and N.\ Uekusa,
%\reftitle{H parity and the stable Higgs boson in the $SO(5)\times U(1)$ gauge-Higgs unification}
\jnl{\PRD}{82}{115024}{2011}.

\bibitem{FHHOS2013}
S.~Funatsu, H.~Hatanaka, Y.~Hosotani, Y.~Orikasa,  and T.~Shimotani,
%\reftitle{Novel universality and Higgs decay $H \go \gamma \gamma, gg$ 
%in the $SO(5) \times U(1)$ gauge-Higgs unification}
\jnl{\PLB}{722}{94}{2013}.

\bibitem{LHCsignalsDM}
S.~Funatsu, H.~Hatanaka, Y.~Hosotani, Y.~Orikasa and T.~Shimotani,
%\reftitle{LHC signals of the $SO(5)\times U(1)$ gauge-Higgs unification}
\jnl{\PRD}{89}{095019}{2014}; 
%\bibitem{GH-DM}
%S.~Funatsu, H.~Hatanaka, Y.~Hosotani, Y.~Orikasa and T.~Shimotani,
%\reftitle{Dark matter in the $SO(5)\times U(1)$ gauge-Higgs unification}
\jnl{\PTEP}{2014}{113B01}{2014}.
%arXiv:1407.3574 [hep-ph].

\bibitem{HZgamma}
S.~Funatsu, H.~Hatanaka and Y.~Hosotani, 
%\reftitle{$H \to Z\gamma$ in the gauge-Higgs unification}
arXiv: 1510.06550 [hep-ph].


\bibitem{Burdman2003}
G.\ Burdman and Y.\ Nomura,
%\reftitle{Unification of Higgs and gauge fields in five dimensions}
\jnl{\NPB}{656}{3}{2003}.

\bibitem{HHKY2004}
N.\ Haba, Y.\ Hosotani, Y.\ Kawamura and T.\ Yamashita,
%\reftitle{Dynamical symmetry breaking in gauge Higgs unification on orbifold}
\jnl{\PRD}{70}{015010}{2004}.

\bibitem{Lim2007}
C.S. Lim and N.\ Maru,
%\reftitle{Towards a realistic grand gauge-Higgs unification}
\jnl{\PLB}{653}{320}{2007}.


\bibitem{Kojima2011}
K.~Kojima, K.~Takenaga and T.~Yamashita,
%\reftitle{Grand gauge-Higgs unification}
\jnl{\PRD}{84}{051701(R)}{2011}.


\bibitem{Serra2011}
M.\ Frigerio, J.\ Serra and A.\ Varagnolo,
%\reftitle{Composite GUTs: models and expectations at the LHC}
\jnl{\JHEP}{1106}{029}{2011}.


\bibitem{HY2015}
Y.\ Hosotani and N.\ Yamatsu, 
%\reftitle{Gauge-Higgs grand unification}
arXiv: 1504.03817 [hep-ph].  To appear in {\it PTEP.}




\end{thebibliography}
\end{document}